\documentclass[twocolumn,showpacs,preprintnumbers,amsmath,amssymb,prb,aps]{revtex4}
\usepackage{epsfig}
\usepackage{graphicx}
\usepackage{dcolumn}
\usepackage{bm}
\begin{document}
\title{Investigation of unoccupied electronic states of LaCoO$_3$ and PrCoO$_3$
using inverse photoemission spectroscopy and GGA\,+\,$U$ calculations}
\author{S. K. Pandey,$^1$ Ashwani Kumar,$^2$ S. Banik,$^1$ A. K. Shukla,$^1$
S. R. Barman$^1$ and A. V. Pimpale$^1$}
\affiliation{$^1$UGC-DAE Consortium for Scientific Research,
University Campus, Khandwa Road, Indore 452 017, India.\\
$^2$Department of Physics, Institute of Science and Laboratory
Education, IPS Academy, Indore 452 012, India.}

\date{\today}

\begin{abstract}
The unoccupied electronic states of LaCoO$_3$ and PrCoO$_3$ are
studied using room temperature inverse photoemission spectroscopy
and \emph{ab initio} GGA+$\emph{U}$ band structure calculations. A
fairly good agreement between experiment and theory is obtained. The
intensity of the peak just above the Fermi-level is found to be very
much sensitive to the hybridization of Co 3$d$ and O 2$p$ orbitals.
Moreover, the band just above the Fermi-level is of Co 3$d$
character with little contribution from O 2$p$ states.

\end{abstract}

 \maketitle

Perovskite type transition metal oxides with general formula
ABO$_{3}$ (A $\equiv$ rare-earth ions and B $\equiv$ transition
metal ions) have been of much interest for more than fifty
years.\cite{imada} Recently, research activities have been
intensified in this class of materials due to the emergence of
exotic properties like charge-disproportionation, charge ordering,
orbital ordering, phase separation, colossal magneto resistance
\emph{etc}. \cite{imada} The interplay between the on-site and
inter-site Coulomb interaction, the charge transfer energy, the
hybridization strength between the cation 3$d$ and oxygen 2$p$
states and the crystal field splitting for the $d^{n}p^{m}$
configuration of the BO$_{6}$ octahedron control the ground state
electronic structure and magnetic and transport properties of
these perovskites.

Cobaltates with general formula ACoO$_{3}$ forms an interesting
class of compounds in the perovskite family. The ground states of
these compounds are believed to be nonmagnetic (spin S = 0)
insulator having Co$^{3+}$ ion in the low spin configuration with
fully filled Co $t_{2g}$ orbitals. These compounds show insulator to
metal and non-magnetic to paramagnetic transitions with increase in
temperature.\cite{heikes,bhide,yamaguchi1,yamaguchi2,tsubouchi} It
is believed that such transition occurs due to thermally driven spin
state transition of Co$^{3+}$ ion.\cite{goodenough,raccah,korotin}
To understand the electronic transport in these compounds the
understanding of both the occupied and the unoccupied electronic
states is desirable. Most of the work found in literature deals only
with occupied part of the electronic states.
\cite{lam,barman,saitoh1,pandey3,saitoh2,pandey1,pandeyPCO}  There
are very few reports in the literature dealing with the unoccupied
electronic states and most of them are based on \emph{x}-ray
absorption studies of LaCoO$_{3}$.\cite{abbate,chainani,pandeyxanes}

Here we study the unoccupied electronic states of LaCoO$_{3}$ and
PrCoO$_{3}$ using inverse photoemission spectroscopy (IPES) and show
how the unoccupied electronic states is affected when La is replaced
by Pr. The replacement of La by Pr has two fold effect: (i) change
in chemical pressure due to changed ionic radius and (ii) change in
the electronic occupancy of 4$f$ states. The change in chemical
pressure may affect the crystal structure of PrCoO$_{3}$. The powder
diffraction work has shown the crystal structure of LaCoO$_{3}$ and
PrCoO$_{3}$ as rhombohedral and orthorhombic, respectively.
\cite{tsubouchi,radaelli} The Pr$^{3+}$ ion contains two electrons
in the 4$f$ states. This would contribute to the valence band and
expected to affect the unoccupied electronic structure of the
compound.

In the present work, we investigate the room temperature unoccupied
electronic structure of LaCoO$_{3}$ and PrCoO$_3$ compounds using
IPES and {\em ab initio} band structure calculation. The comparison
of experimental spectra with the convoluted total density of states
indicates that GGA+$\emph{U}$ calculation is sufficient for
understanding all experimentally observed features in the spectrum.
On going from LaCoO$_{3}$ to PrCoO$_{3}$ intensity of peak just
above the Fermi-level decreases for PrCoO$_{3}$ due to decrease in
the hybridization of Co 3$d$ and O 2$p$ orbitals in this compound in
comparison to LaCoO$_{3}$. The band just above the Fermi-level  is
of Co 3$d$ character.

LaCoO$_{3}$ and PrCoO$_{3}$ were prepared in the polycrystalline
form by combustion method.\cite{pandey2} Nitrates of La, Pr and Co
were taken in appropriate amount and mixed in double distilled
water. In this mixture, 2 moles of glycine per 1 mole of metal
cation was added and stirred until all the compounds dissolved in
the water. The resulting solution was heated slowly at temperature
around 200$^{0}$C till all the water evaporated. The precursor
thus formed catches fire on its own making the powder of the
desired compound. The hard pellets of this powder were formed and
heated at 1200$^{0}$C for one day. The samples were characterized
by $x$-ray powder diffraction technique. The powder diffraction
data did not show any impurity peak; all the peaks of LaCoO$_{3}$
and PrCoO$_{3}$ were well fitted with rhombohedral (space group
\emph{R$\overline{3}$c}) and orthorhombic (space group
\emph{Pbnm}) phases, respectively, using Rietveld refinement
technique. The lattice parameters obtained from the fitting match
well with those reported in the literature.

The inverse photoemission spectroscopy experiments on LaCoO$_{3}$
and PrCoO$_{3}$ compounds were performed under ultrahigh vacuum at a
base pressure of 6 $\times$ 10$^{-11}$ mbar. The samples were
mounted in the form of a compressed hard pellet and they were
scraped uniformly by diamond file to obtained clean surface. An
electrostatically focused electron gun of Stoffel Johnson design and
an acetone gas filled photon detector with a CaF$_{2}$ window were
used for the experiments.\cite{funnemann,banik} The experiments
were carried out in the isochromat mode where the kinetic energy of
the incident electrons were varied at 0.05 eV steps and photons of
fixed energy (9.9 eV) were detected with an overall resolution of
0.55 eV.\cite{banik} Fermi-level was aligned by recording the IPES
spectrum of {\it in situ} cleaned silver foil.

The GGA+$U$ spin-polarized density of states (DOS) calculations of
LaCoO$_{3}$ were carried out using LMTART 6.61.\cite{savrasov} For
calculating charge density, full-potential linearized Muffin-Tin
orbital method working in plane wave representation was employed. In
the calculation, we have used the Muffin-Tin radii of 3.509, 2.001,
and 1.674 a.u. for La, Co and O, respectively. The charge density
and effective potential were expanded in spherical harmonics up to
$l$= 6 inside the sphere and in a Fourier series in the interstitial
region. The initial basis set included 6$s$, 6$p$, 5$d$, and 4$f$
valence, and 5$s$ and 5$p$ semicore orbitals of La; 4$s$, 4$p$, and
3$d$ valence, and 3$p$ semicore orbitals of Co, and 2$s$ and 2$p$
valence orbitals of O. The exchange correlation functional of the
density functional theory was taken after Vosko {\em et
al.}\cite{vosko} and generalized gradient approximation (GGA) was
implemented using Perdew {\em et al.} prescription.\cite{perdew} In
the GGA+$U$ calculations the Hubbard $U$ and exchange $J$ are
considered as parameters. We have taken $U$=\,3.5 eV and $J$=\,1.0
eV for Co 3$d$ electrons. The values of $U$ and $J$ for 3$d$
electrons are consistent with our previous
studies.\cite{pandey1,pandey3} Self-consistency was achieved by
demanding the convergence of the total energy to be smaller than
10$^{-5}$ Ryd/cell. (6, 6, 6) divisions of the Brillouin zone along
three directions for the tetrahedron integration were used to
calculate the density of states.

The experimental spectra of LaCoO$_{3}$ and PrCoO$_{3}$ are
plotted in Fig. 1.  The normalized spectra of these compounds in
the region closer to Fermi-level are given in the inset of the
figure. Three features are clearly seen in the spectrum
corresponding to LaCoO$_{3}$ denoted by A, B and C. Peak B is not
visible in the PrCoO$_{3}$ spectrum. It may be noted that the
first peak A is somewhat more asymmetric for lanthanum compound as
compared to praseodymium one. This may be an indication of some
additional structure around this energy in LaCoO$_{3}$. The
feature B is not seen in the bremsstrahlung isochromat
spectroscopy (BIS) data of LaCoO$_{3}$ of Chainani {\it et
al.}\cite{chainani} This may be due to different cross-section and
worse resolution of the BIS data. It is clear from the inset that
there is little intensity near the Fermi-level taken as energy
origin indicating the insulating nature of the compounds. One may
note the rise in the intensity above the Fermi-level, which is
relatively sharper for LaCoO$_{3}$ in comparison to PrCoO$_{3}$.
Also the height of peak A corresponding to PrCoO$_{3}$ is less in
comparison to that in LaCoO$_{3}$.

The experimental and calculated spectra of LaCoO$_{3}$ are plotted
in Fig. 2. The calculated spectrum also contains three features
denoted by arrows corresponding to experimentally observed
features A, B and C. In addition a structure is seen around 3 eV,
which may account for the asymmetry of the peak A in the
experimental data. There is some deviation in the energy positions
and the corresponding intensities of the features. Such deviation
may be due to approximations involved in the calculations. It
should be mentioned here that our calculations of IPES is based
only on total DOS and matrix element as well as inelastic
background are not considered. Keeping this in mind, the agreement
between experiment and calculation as far as features are
concerned is reasonably satisfactory.

To identify the contribution of different partial DOS in different
features, we have plotted the Co 3$d$, O 2$p$, La 5$d$ and La 4$f$
partial DOS in Fig. 3 as these DOS contribute most in the energy
regions of interest. It is evident from the panels 1 and 2 that in
feature A, Co 3$d$ and O 2$p$ states contribute. There is a
substantial rise in Co 3$d$ DOS just above the Fermi level
indicating that the region just above it has only Co 3$d$
character. Feature B mainly arises from electronic transition to
mixed La 5$d$ and La 4$f$ states. The feature C can be attributed
to mixed O 2$p$, La 5$d$ and La 4$f$ states. It is evident from
panels 2, 3 and 4 that the La 4$f$ partial DOS is much larger than
the O 2$p$ and La 5$d$ partial DOS. Therefore, any change in the
number of electrons in the 4$f$ orbital will have direct effect on
the intensity of features B and C. This is clearly evident from
figure 1 where the intensity of these features decreases for
PrCoO$_{3}$ in comparison to LaCoO$_{3}$. In PrCoO$_{3}$ there are
two electrons in the occupied 4$f$ band, whereas in LaCoO$_{3}$
the 4$f$ band is fully unoccupied. Thus one expects a decrease in
the intensity of features having 4$f$ character (in the case of
PrCoO$_{3}$) and this is clearly seen in the data. The reduced
asymmetry and intensity under peak A for PrCoO$_{3}$ can be seen
to be due to unoccupied O 2$p$ DOS. It may be noted that the
overlap of Co 3$d$ and O 2$p$ orbitals is some what reduced for
the PrCoO$_{3}$ in comparison to LaCoO$_{3}$ due to increased
average Co-O bond length as revealed by our extended \emph{x}-ray
absorption fine structure (EXAFS) studies.\cite{pandeyexafs} The
decrease in overlap will reduce the transfer of electron from O
2$p$ to Co 3$d$ orbital and as a consequence decrease in
unoccupied O 2$p$ band. The interpretation is in line with the
plotted partial DOS in panel 2 of figure 3 where O 2$p$ DOS
contributing to these features is 1 eV above the Fermi-level from
where the intensity started decreasing for PrCoO$_{3}$, as is
evident from Fig. 1. This behaviour of intensity clearly indicates
that the band just above the Fermi-level is of $d$ character
arising from Co 3$d$ partial DOS. This result is in contrast with
configuration interaction calculations of Saitoh {\it et al.}
\cite{saitoh4} giving O 2$p$ character of band just above the
Fermi-level for LaCoO$_{3}$.

In conclusion, we have carried out room temperature inverse
photoemission spectroscopy (IPES) measurements on LaCoO$_{3}$ and
PrCoO$_{3}$ compounds. The IPES data are analyzed by using density
of states obtained from GGA+$\emph{U}$ calculation. It is noted that
such calculations might suffice to understand the features observed
in the experimental spectra up to 12 eV above the Fermi-level. The
hybridization of Co 3$d$ and O 2$p$ orbitals is found to be crucial
for understanding the change in intensity of the peak just above the
Fermi-level. The band just above it is of Co 3$d$ character with
little contribution from O 2$p$ states.

P. Chaddah, A. Gupta and K. Horn are thanked for constant encouragement.
Financial support from Department of Science and Technology,
Government of India through project No. SP/S2/M-06/99 and Ramanna fellowship research grant
is gratefully acknowledged.

\pagebreak

\section{Figure Captions:}
Fig. 1: (color online)  Inverse photoemission spectra of LaCoO$_{3}$
and PrCoO$_{3}$. Inset shows the normalized spectra near the
Fermi-level of these compounds. The energy origin is taken at the
Fermi-level.

Fig. 2: (color online) The experimental and calculated inverse
photoemission spectra of LaCoO$_{3}$.

Fig. 3: Partial density of states of Co 3$d$, O 2$p$, La 5$d$ and La
4$f$ characters.

\end{document}